\documentclass[%
 reprint,
 amsmath,amssymb,
prb,
]{revtex4-1}

\usepackage{graphicx}
\usepackage{dcolumn}
\usepackage{bm}

\begin{document}

\preprint{APS/123-QED}

\title{Plasmonic Aerosols}

\author{Jeffrey Geldmeier}

 \author{Paul Johns}

  \author{Nicholas J. Greybush}
 
   \author{Jawad Naciri}
\author{Jake Fontana}%
 \email{Corresponding author: jake.fontana@nrl.navy.mil}
\affiliation{%
 U.S. Naval Research Laboratory\\
 4555 Overlook Ave. SW, Washington, D.C. 20375
}%

\date{\today}%

\begin{abstract}
Plasmonic nanoparticles resonantly couple to and confine light below the diffraction limit. This mechanism has enabled a modern renaissance in optical materials, with potential applications ranging from sensing and circuitry to renewable energies and medicines. However, these plasmonic materials are typically constrained to dilute liquids or solid two-dimensional surfaces, thereby limiting their possibilities. Here, we experimentally demonstrate a plasmonic aerosol by transitioning liquid suspensions of gold nanorods into the gas phase and simultaneously measuring their optical spectra. By measuring and modeling the evolution of the longitudinal absorbance peak of the nanorods from the liquid to the gas phase, we find that the aerosols are optically homogeneous and thermodynamically stable. We show that by tailoring the aspect ratio of the nanorods, the aerosol absorbance peak is tunable from visible to midwave infrared wavelengths. We find the sensitivity of the absorbance peak wavelength to changes in the refractive index of the gas depends linearly on the aspect ratio and can be estimated from the geometric and material properties of the nanorod. For high aspect ratio nanorods the sensitivity becomes extremely large, which  may  be  useful  in  aiding  geoengineering challenges. This work establishes plasmonic aerosols, potentially enabling exciting opportunities for fundamental and applied research.

\begin{description}

\item[PACS numbers]
71.45.Gm, 42.68.Jg, 78.67.Qa, 64.70.Nd
\end{description}
\end{abstract}

\pacs{Valid PACS appear here}
\maketitle


\section{\label{sec:level1}Introduction}

The relationship between aerosol particles and cloud systems is a poorly understood nonlinear process and is the largest uncertainty to accurately predicting climate and extreme weather events.\cite{IPOC2013, Rosenfeld2018} Aerosol particles serve as nucleation sites for water molecules to condense into droplets that can then form into clouds. Recent work posited that aerosol particles from the exhaust of ships enhanced the intensity and electrification of storms, showing that the density of lightning strikes doubled over shipping lanes.\cite{Thornton2017} Moreover, ultrafine aerosol particles (diameter $< 50~nm$), once thought to be too small to influence cloud formation, have recently been shown to significantly intensify the convective strength of cloud systems,\cite{Rosenfeld2018} indicating that nanoparticle aerosols may also be used for geoengineering applications.\cite{Keith2000, Keith2010,Palmer1980,Palmer1983, Leung1985, Besteiro2017, Besteiro2018}

The influence of nanoparticle aerosols on cloud formation is extremely complex and hard to disentangle, and a significant need exists to experimentally model these systems in controlled environments to carefully examine the nanoscale mechanisms governing these macro-scale processes. Aerosols composed of micrometer-sized particles have been thoroughly investigated for decades.\cite{Bohren1983} However, the experimental aerosolization and optical detection of nanoparticle aerosols is a longstanding challenge due to factors such as aggregation upon the liquid-gas phase transition, relatively dilute concentrations, or small light-matter coupling.\cite{Koman2018}

Plasmonic nanoparticles are promising candidates for benchtop aerosol studies. They couple strongly to light, leading to the capability to optically detect them in dilute concentrations, and they are also sensitive to changes in their surrounding environment. A simple harmonic oscillator model can be used to describe the behavior of the plasmonic nanoparticles in an optical field.\cite{Bohren1983}  From this model, the imaginary electric susceptibility of a plasmonic nanoparticle is $\chi''=\beta\omega_{p}^{2} \omega /((L\omega_{p}^{2}-\omega^{2})^{2}+\beta^{2} \omega^{2})$, where $\beta$ is the damping constant, $L$ is the depolarization factor, $\omega_{p}$ is the plasma frequency, and $\omega$ is the frequency of the incident light. The imaginary susceptibility, and consequently the absorption, is a maximum at resonance, $\omega=\sqrt{L}\omega_{p}$, yielding, $\chi''_{max}=\omega_{p}/(\beta \sqrt{L})$. Therefore, a pragmatic nanoparticle to maximize the absorption is a gold nanorod\cite{Miller2014,Miller2016,Kravets2008} due to its large $\omega_{p}$, small $L$ (along the long axis of the nanorod), and mature chemical-based fabrication. 

The nanorods will be thermodynamically stable in the gas state when the gravitational force, $\Delta\rho V g $, is less than the stabilizing thermal forces, $k_{B}T/l$, where $\Delta\rho$ is the density difference between the nanorod and gas, $V=\frac{4}{3}\pi r^{2}l$ is the volume of the nanorod, $l$ is the length and $r$ is the radius of the nanorod, $k_{B}$ is the Boltzmann constant and $T$ is the absolute temperature.  Accordingly, if the length of the gold nanorods is smaller than $\sqrt{(3k_{B}T/4\pi\Delta\rho g r^{2})} \approx \mu m$, then they will remain suspended in the gas state.  Gold is also an inert metal, making it bio-compatible and environmentally friendly.  Additionally, recent gram-scale, colloidal gold nanorod synthesis breakthroughs have now made these materials accessible in large quantities.\cite{Park2017}

Here, we solve a decades-old problem of simultaneously aerosolizing and measuring the optical response of plasmonic nanoparticles in the gas phase, thereby uniting the fields of plasmonics and aerosols.  We show that the aerosols are optically homogeneous, thermodynamically stable, with wide wavelength tunability, and extremely high sensitivities to their environment that may be useful in aiding geoengineering challenges. We anticipate plasmonic aerosols will open up broad and innovative approaches to understand the underlying physics of inaccessible climatology, astronomy, petroleum and medical environments.  In the context of vacuum microelectronics,\cite{Stoner2012,Srisonphan2012,Jones2017} if plasmonic aerosols are encapsulated into micron-sized elements and gated using external electric fields, then the electro-optic properties of the element may be reconfigurable by controlling the orientational order of the nanorods.\cite{Fontana2016, Etcheverry2017} These materials may also be useful for nonlinear optics,\cite{Palmer1983, Palmer1980, Leung1985} nanojet printing,\cite{Gupta2018} molecular diagnostics,\cite{Phan2018} or nanomedicines.\cite{Raliya2017}

\section{\label{sec:level1}Results and Discussion}

In Fig. 1(a), we show a schematic for the transition of aqueous suspensions of gold nanorods from the liquid phase into the gas phase while simultaneously measuring their optical response (see Supplemental Material for details). The gold nanorods were synthesized using wet-seed mediated synthesis approaches that enabled tuning of the aspect ratio (length, $l$, to diameter, $d$) from $1.5$ to $38$.\cite{Park2017,Takenaka2009,Zubarev2008}  Transmission electron microscopy (TEM) was used to measure the polydispersity of the nanorods, yielding less than $10\%$ for aspect ratios smaller than $20$ and $25\%$ for larger aspect ratios.  To aerosolize the nanorods, a Venturi tube was used to drive high-velocity air over a reservoir of gold nanorods in a liquid suspension, pulling the suspension into the airstream.  Upon exiting the tube, the liquid suspension breaks apart into an aerosol containing liquid droplets (diameter $\sim300~nm$) with embedded nanorods. The droplets then enter a dehumidifier chamber that evaporates the remaining water from the nanorods, thereby creating a suspension of dry nanorods in the gas phase. To measure the \textit{in~situ} absorbance spectra of the gold nanorods in the gas phase, a $10~m$ optical path length Herriott cell was placed at the exit of the dehumidifier. The Herriott cell was integrated into a Fourier transform infrared (FTIR) spectrometer, enabling the optical signatures of the plasmonic aerosols to be measured from $0.6-3~\mu m$ or $1-10~\mu m$ in a continuous manner.

\begin{figure}[htbp]
\centering\includegraphics[width=8.5cm]{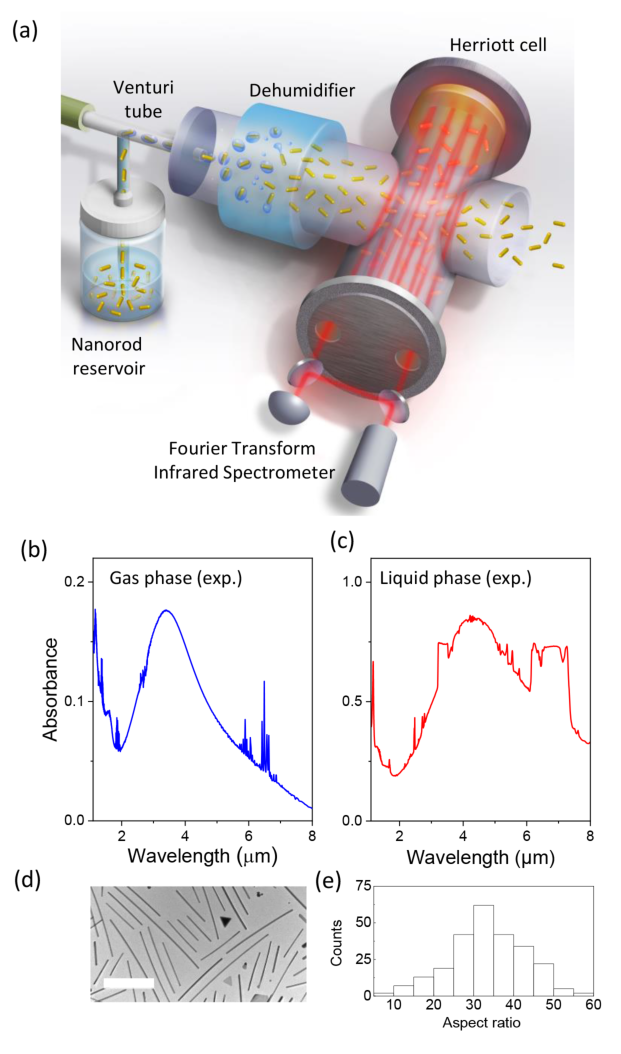}
\vspace{-0.5cm}
\caption{Schematic of the experimental apparatus to aerosolize and optically measure gold nanorods in the gas phase \textit{in~situ} is shown in (a). The absorbance spectra of high aspect ratio gold nanorods in the gas3402, (b), and liquid phase, (c). A representative TEM image is shown in (d) with the corresponding aspect ratio statistics depicted in (e). The scale bar in (d) is $500~nm$.}
\label{1}
\end{figure}

\begin{figure}[htbp]
\centering\includegraphics[width=8.5cm]{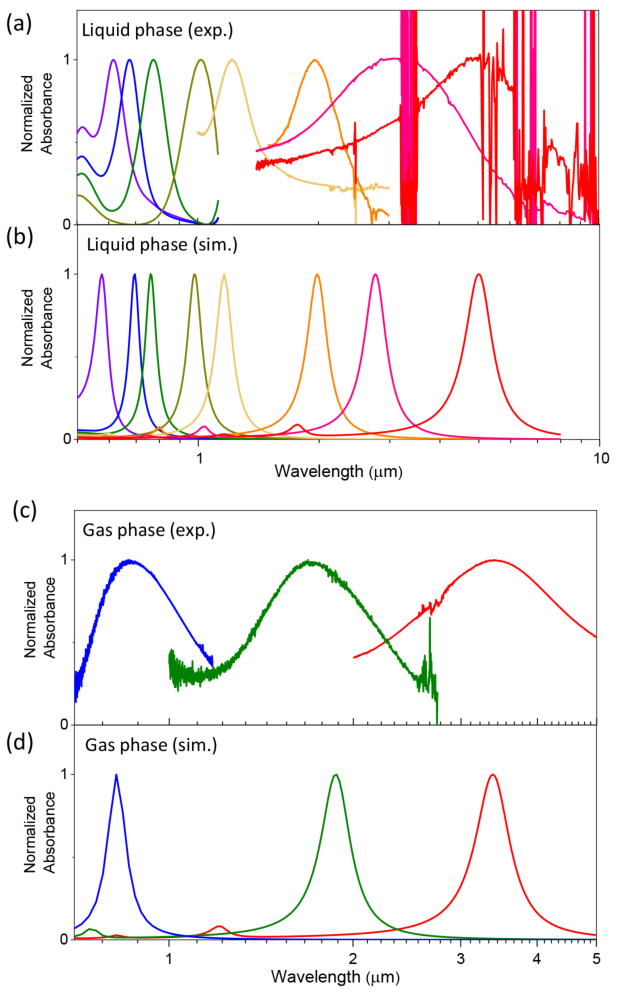}
\vspace{-0.5cm}
\caption{Experimental spectra of gold nanorods with aspect ratios ranging from $1.5$, $2.5$, $3$, $4.5$ in water to $10$, $15$, $30$ in toluene, (a), and $5$, $15$, $30$ in air (c). Three-dimensional finite element simulation spectra in (b) and (d) matching the experimental parameters and phases in (a) and (c).}
\label{1}
\end{figure}

The absorbance spectra of the gold nanorods in the gas phase is shown in Fig. 1(b), revealing a well defined absorption peak at $3.3~\mu m$. The relatively large effective Q-factor $(\lambda_{0}/FWHM)$ of $1.3$ in Fig. 1(b) suggests that the nanorods are not flocculated or aggregated upon transitioning from the liquid to the gas phase. The magnitude of the absorbance peak was constant while the measurements were performed (average of 16 individual spectra) demonstrating the aerosols are temporally stable.  The constant absorbance peak also implies the aerosols are optically homogeneous, which is expected since the nanorods are generally much smaller than the wavelength of the light ($\sim\lambda_{0}/10$). The sharp absorbance peaks around $6~\mu m$ are due to the $O-H$ molecular vibrations from trace amounts of water vapor. The apparent increase in absorbance near $1~\mu m$ may be attributed to these wavelengths being near the wavelength detection limit for the specific FTIR detector-beamsplitter combination, although triangular platelet by-products from the high aspect nanorod synthesis may have resonances in this region as well.  The density of nanorods in the Herriott cell is estimated to be $\rho=A/\sigma x\approx 10^{11} ~NR/m^3$, where $A=0.18$ is the peak absorbance at $3.3~\mu m$, $\sigma=9.4\times 10^{-14} ~m^2/NR$ is the extinction cross section of the nanorod, which was retrieved from three-dimensional finite element simulations (COMSOL Multiphysics, v. 5.3a) and $x=10~m$ is the path length of light through the Herriott cell.

Due to water's prohibitively large absorption bands beyond $1.2~\mu m$, the nanorods were phase transferred from water into toluene suspensions\cite{Fontana2016} to enable the absorbance spectra in the infrared region to be measured in the liquid phase, as seen in Fig. 1(c). While molecular vibrations from the aromatic ring of toluene are still significant throughout the spectrum, leading to detector saturation in some wavelength bands, the absorbance peak from the nanorods is apparent. TEM images, as shown in Fig. 1(d), were used to determine an average aspect ratio of $32.3$ ($l=646\pm 156~nm$, $d=20\pm 4~nm$) as shown in Fig. 1(e).

The aspect ratio of the gold nanorods was varied from $1.5$, $2.5$, $3$, and $4.5$ in water to $10$, $15$, and $30$ in toluene in Fig. 2(a), demonstrating nearly a decade in experimental wavelength tunability from $0.6$ to $5~\mu m$ in the liquid phase. The effective Q-factor in the liquid phase varied from $5$ to $1$ for aspect ratios ranging from $1.5$ to $30$, respectively.

\begin{figure}[htbp]
\centering\includegraphics[width=8.5cm]{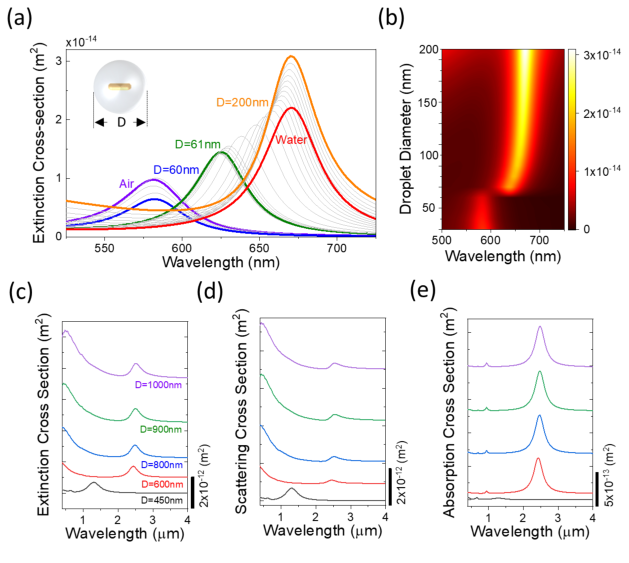}
\vspace{-0.5cm}
\caption{Extinction spectra (a) and color map (b) of a nanorod surrounded by water and air, as the diameter of a water droplet, $D$, encasing the nanorod decreases. The extinction, scattering, and absorption cross-section of a gold nanorod  as the diameter of the water droplet decreases are presented in (c), (d) and (e).}
\label{1}
\end{figure}

The optical response of gold nanorods in the gas phase with aspect ratios of $5$, $15$, and $30$ are shown in Fig. 2(c), spanning over $2.5~\mu m$ in wavelength tunability. The effective Q-factor in the gas phase varied from $2.4$ to $1.3$ for aspect ratios of $5$ to $30$, respectively.

Three-dimensional finite element simulations in Fig. 2(b) and Fig. 2(d) were used to replicate the experimental isotropic spectra of the nanorods in the liquid and gas states.  The simulated longitudinal absorbance peak wavelengths are in good agreement with the experimental values as a function of aspect ratio and solvent index.  The effective Q-factors of the longitudinal absorbance peaks from the simulated spectra are larger than for the experimental spectra due to experimental variations in the aspect ratio of the nanorods.\cite{Eustis2006}  The small blue-shifted absorbance peaks, relative to the longitudinal peaks, are due to the transverse axis of the nanorods.

To understand the spectroscopic evolution of the evaporating water-nanorod droplets, further simulations were carried out in Fig. 3.  Initially, a nanorod ($l=60~nm$, $d=25~nm$) was surrounded with water, yielding an extinction peak wavelength at $670~nm$ (red curve), Fig. 3(a). The nanorod was then embedded into a water droplet of diameter $D=200~nm$ and surrounded by air (orange curve). The magnitude of the peak is increased by $\sim 1/3$ compared to a homogenous water medium due to the increased scattering contribution to the extinction from the water droplet surface. As the diameter of the water droplet decreased to $61~nm$, mimicking evaporation, the peak wavelength blue shifted to $624~nm$ (green curve). The extinction peak then drastically shifted to $582~nm$ when $D=l=60~nm$, as seen in Fig. 3(b), due to the ends of the nanorod being exposed to air. For droplet diameters less than the length of the nanorod the extinction peak remains constant, only slightly increasing in magnitude once all the water is removed.

Fig. 3(c-e) shows the extinction, scattering, and absorption cross sections of a high aspect ratio gold nanorod $(l=450~nm, d=20~nm)$ as the diameter of the water droplet decreases from $D=1,000~nm$ to $D=l=450~nm$. The most striking feature as the water droplet evaporates is the decrease in the visible wavelength scattering. The albedo, which is the ratio of the scattering cross-section to the extinction cross-section, is $0.066$ for the $l=60~nm$ nanorods and $0.366$ for the $l=450~nm$ nanorods.

Transitioning the nanorods from the liquid to the gas state results in large shifts in the absorbance peak wavelength, as shown in Fig. 2. By monitoring the peak shifts and comparing the measurements with simulations, we can accurately determine the local refractive index of the medium surrounding the nanorods and therefore infer the state of the nanorod suspension, e.g. water, toluene, or air.  Moreover, the effective Q-factor is a good metric to ensure the nanorods are not aggregated, as discussed with Fig. 2.

To validate that the nanorods are dispersed in air, we carried out experiments and simulations measuring the absorbance peak wavelength at various aspect ratios of the gold nanorods: $7.5$, $13$, $16.8$, $33$, and $38$ in toluene and air (Fig. 4(a)).

The simulations show that the absorbance peak wavelength depends linearly on the aspect ratio of the nanorods, $\lambda_{toluene}=0.135(l/d)+0.422$ and $\lambda_{air}=0.0883(l/d)+0.359$, and the peak red-shifts as the refractive index of the host medium increases.\cite{Fontana2017}  

We find the experimental data agree well with the simulation data, within experimental uncertainty, and that the absorbance peak is proportional to the aspect ratio. Furthermore, the experimental data for the nanorods in air agree well with the simulation predictions for their peak wavelengths, implying the nanorods are homogeneously dispersed as an aerosol.

To further confirm that the nanorods are uniformly suspended in air, the longitudinal absorption peak wavelength can be related to the refractive index of the host medium, $n_{m}$,\cite{Yang2005} by:

\begin{equation}\lambda=\lambda_{p}\sqrt {1+(1-L_{\parallel})(n_{s}^{2}-n_{m}^{2})f+\Bigg(\frac{1}{ L_{\parallel}}-1\Bigg) n_{m}^{2}}
\end{equation}
where $\lambda_{p}$ is the plasma wavelength of gold, $n_{s}$ is the refractive index of the ligand shell coating the nanorods, and $f$ is the ellipsoidal volume fraction of the inner nanorod to the outer ligand shell. The depolarization factor of the long axis of the nanorod is $L_{\parallel}=((1-\epsilon^{2})/\epsilon)((1/2\epsilon)(ln(1-\epsilon)/(1+\epsilon))-1)$, where $\epsilon=\sqrt{1-(l/d)^{2}}$.

If the nanorods are very long $(1/L_{\parallel}\gg 1)$ and there is no ligand shell $(n_{s}=n_{m})$, then Eq. 1 can be differentiated with respect to $n_{m}$ and then series expanded about $L_{\parallel}$ to approximate the sensitivity, 

\begin{equation}
\frac{\partial \lambda}{\partial n_{m}} \approx \frac{\lambda_{p}}{\sqrt{L_{\parallel}}}
\end{equation}

This result implies that if the geometric ($L_{\parallel}$) and material ($\lambda_{p}$) properties of the nanorod are known, the shift in the absorption peak wavelength can be estimated as the host medium surrounding the nanorods is varied.

\begin{figure}[htbp]
\centering\includegraphics[width=8.5cm]{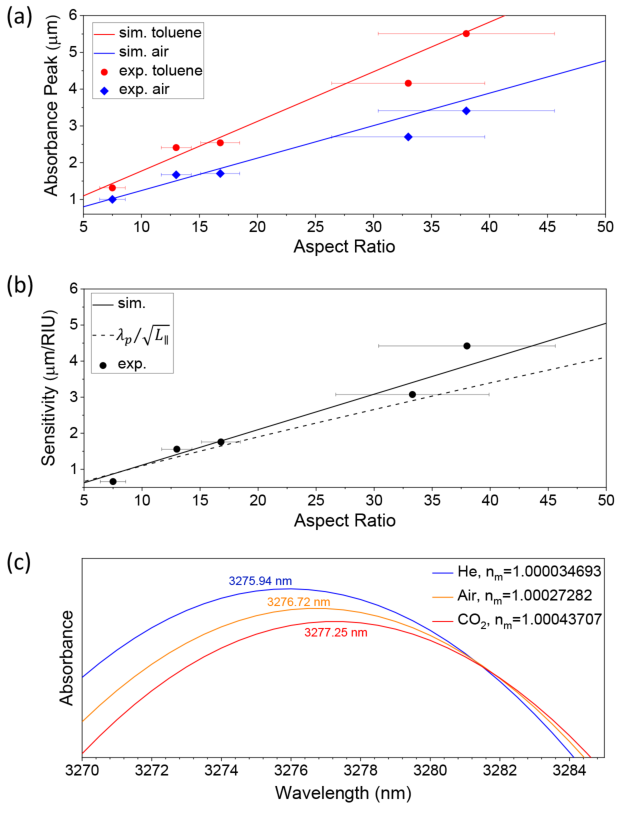}
\vspace{-0.5cm}
\caption{Evolution of the absorbance peak wavelength and sensitivity as the nanorod aspect ratio and host refractive indexes are varied are shown in (a) and (b), respectively.  The simulated absorbance spectra for plasmonic aerosols as a function of the host gas is shown in (c).}
\label{1}
\end{figure}

In Fig. 4(b), we show the sensitivity also depends linearly on the aspect ratio, $\partial \lambda/\partial n_{m}=0.098(l/d)+0.133$, for the data retrieved from the simulations (solid line). This result is in good agreement with the experimental data. For smaller aspect ratios, the relationship in Eq. 2 is confirmed (dashed line). For larger aspect ratios, the predictions of Eq. 2 begin to deviate from the simulation data (e.g. $18\%$ at $\textrm{aspect ratio}=45$), showing the limitations of the simple relationship.  Eq. 2 provides a straightforward means to predict the sensitivity, and the good agreement with experimental and simulation data further supports the nanorods being thermodynamically stabilized in the gas phase.

The absorbance peak wavelength shifts are on the order of several microns per refractive index unit (RIU) in Fig. 4(b). This large sensitivity implies that very small changes in the host medium can be detected, suggesting that plasmonic aerosols may be good candidates to accurately probe and model remote environments such as atmospheric systems at benchtop size scales. To investigate this possibility, gold nanorods with an aspect ratio of $30$ were simulated in He, Air, and CO$_{2}$ gaseous environments (Fig. 4(c)). The absorbance peak wavelength shifted from $3,275.94~nm$ for He to $3,277.25~nm$ for CO$_{2}$, showing that changes of $\Delta n_{m}\approx 10^{-4}$ may be detectable at atmospheric transmission window wavelengths.\cite{Neubrech2008,Raman2011} 

\section{\label{sec:level1}Conclusion}

In summary, we demonstrated for the first time the aerosolization of gold nanorods from concentrated liquid suspensions, while simultaneously measuring their optical spectra at bench-top scales. The plasmonic aerosol absorption peaks are sharp and well-defined with effective quality factors as large as $2.4$. We show that by controlling the aspect ratio of the nanorods, the aerosol absorption peaks are broadly tunable over $2,500~nm$ from visible to mid-wave infrared wavelengths. We find that the sensitivity of the longitudinal absorption peak wavelength to the refractive index of the host medium depends linearly on the nanorod aspect ratio and can be estimated from the geometric and material properties of the nanorod. Utilizing this sensitivity dependence, we also show that minute changes of the host refractive index of $10^{-4}$ may be detectable, suggesting these materials could be useful for environmental or remote sensing.

\section{Acknowledgement}

We thank Kyoungweon Park, Richard Vaia, Tiffany Sutton, Jerold Bottiger, Brendan Delacy, and Nader Engheta for useful discussions. Jeffrey Geldmeier and Paul Johns thank the American Society for Engineering Education and Nicholas Greybush thanks the National Research Council for postdoctoral fellowships. This material is based upon work supported by the Office of Naval Research under (N0001418WX00122). See Supplemental Material at https://journals.aps.org/prb/.

\bibliography{references}

\begin{thebibliography}{31}%
\makeatletter
\providecommand \@ifxundefined [1]{%
 \@ifx{#1\undefined}
}%
\providecommand \@ifnum [1]{%
 \ifnum #1\expandafter \@firstoftwo
 \else \expandafter \@secondoftwo
 \fi
}%
\providecommand \@ifx [1]{%
 \ifx #1\expandafter \@firstoftwo
 \else \expandafter \@secondoftwo
 \fi
}%
\providecommand \natexlab [1]{#1}%
\providecommand \enquote  [1]{``#1''}%
\providecommand \bibnamefont  [1]{#1}%
\providecommand \bibfnamefont [1]{#1}%
\providecommand \citenamefont [1]{#1}%
\providecommand \href@noop [0]{\@secondoftwo}%
\providecommand \href [0]{\begingroup \@sanitize@url \@href}%
\providecommand \@href[1]{\@@startlink{#1}\@@href}%
\providecommand \@@href[1]{\endgroup#1\@@endlink}%
\providecommand \@sanitize@url [0]{\catcode `\\12\catcode `\$12\catcode
  `\&12\catcode `\#12\catcode `\^12\catcode `\_12\catcode `\%12\relax}%
\providecommand \@@startlink[1]{}%
\providecommand \@@endlink[0]{}%
\providecommand \url  [0]{\begingroup\@sanitize@url \@url }%
\providecommand \@url [1]{\endgroup\@href {#1}{\urlprefix }}%
\providecommand \urlprefix  [0]{URL }%
\providecommand \Eprint [0]{\href }%
\providecommand \doibase [0]{http://dx.doi.org/}%
\providecommand \selectlanguage [0]{\@gobble}%
\providecommand \bibinfo  [0]{\@secondoftwo}%
\providecommand \bibfield  [0]{\@secondoftwo}%
\providecommand \translation [1]{[#1]}%
\providecommand \BibitemOpen [0]{}%
\providecommand \bibitemStop [0]{}%
\providecommand \bibitemNoStop [0]{.\EOS\space}%
\providecommand \EOS [0]{\spacefactor3000\relax}%
\providecommand \BibitemShut  [1]{\csname bibitem#1\endcsname}%
\let\auto@bib@innerbib\@empty
\bibitem [{\citenamefont {Change}(2013)}]{IPOC2013}%
  \BibitemOpen
  \bibfield  {author} {\bibinfo {author} {\bibfnamefont {I.~P. O.~C.}\
  \bibnamefont {Change}},\ }\href@noop {} {\emph {\bibinfo {title} {Climate
  Change 2013: The Physical Science Basis}}}\ (\bibinfo  {publisher} {Cambridge
  University Press},\ \bibinfo {year} {2013})\BibitemShut {NoStop}%
\bibitem [{\citenamefont {Fan}\ \emph {et~al.}(2018)\citenamefont {Fan},
  \citenamefont {Rosenfeld}, \citenamefont {Zhang}, \citenamefont {Giangrande},
  \citenamefont {Li}, \citenamefont {Machado}, \citenamefont {Martin},
  \citenamefont {Yang}, \citenamefont {Wang}, \citenamefont {Artaxo},
  \citenamefont {Barbosa}, \citenamefont {Braga}, \citenamefont {Comstock},
  \citenamefont {Feng}, \citenamefont {Gao}, \citenamefont {Gomes},
  \citenamefont {Mei}, \citenamefont {Pohlker}, \citenamefont {Pohlker},
  \citenamefont {Poschl},\ and\ \citenamefont {de~Souza}}]{Rosenfeld2018}%
  \BibitemOpen
  \bibfield  {author} {\bibinfo {author} {\bibfnamefont {J.}~\bibnamefont
  {Fan}}, \bibinfo {author} {\bibfnamefont {D.}~\bibnamefont {Rosenfeld}},
  \bibinfo {author} {\bibfnamefont {Y.}~\bibnamefont {Zhang}}, \bibinfo
  {author} {\bibfnamefont {S.~E.}\ \bibnamefont {Giangrande}}, \bibinfo
  {author} {\bibfnamefont {Z.}~\bibnamefont {Li}}, \bibinfo {author}
  {\bibfnamefont {L.~A.~T.}\ \bibnamefont {Machado}}, \bibinfo {author}
  {\bibfnamefont {S.~T.}\ \bibnamefont {Martin}}, \bibinfo {author}
  {\bibfnamefont {Y.}~\bibnamefont {Yang}}, \bibinfo {author} {\bibfnamefont
  {J.}~\bibnamefont {Wang}}, \bibinfo {author} {\bibfnamefont {P.}~\bibnamefont
  {Artaxo}}, \bibinfo {author} {\bibfnamefont {H.~M.~J.}\ \bibnamefont
  {Barbosa}}, \bibinfo {author} {\bibfnamefont {R.~C.}\ \bibnamefont {Braga}},
  \bibinfo {author} {\bibfnamefont {J.~M.}\ \bibnamefont {Comstock}}, \bibinfo
  {author} {\bibfnamefont {Z.}~\bibnamefont {Feng}}, \bibinfo {author}
  {\bibfnamefont {W.}~\bibnamefont {Gao}}, \bibinfo {author} {\bibfnamefont
  {H.~B.}\ \bibnamefont {Gomes}}, \bibinfo {author} {\bibfnamefont
  {F.}~\bibnamefont {Mei}}, \bibinfo {author} {\bibfnamefont {C.}~\bibnamefont
  {Pohlker}}, \bibinfo {author} {\bibfnamefont {M.~L.}\ \bibnamefont
  {Pohlker}}, \bibinfo {author} {\bibfnamefont {U.}~\bibnamefont {Poschl}}, \
  and\ \bibinfo {author} {\bibfnamefont {R.~A.~F.}\ \bibnamefont {de~Souza}},\
  }\href {\doibase 10.1126/science.aan8461} {\bibfield  {journal} {\bibinfo
  {journal} {Science}\ }\textbf {\bibinfo {volume} {359}},\ \bibinfo {pages}
  {411} (\bibinfo {year} {2018})}\BibitemShut {NoStop}%
\bibitem [{\citenamefont {Thornton}\ \emph {et~al.}(2017)\citenamefont
  {Thornton}, \citenamefont {Virts}, \citenamefont {Holzworth},\ and\
  \citenamefont {Mitchell}}]{Thornton2017}%
  \BibitemOpen
  \bibfield  {author} {\bibinfo {author} {\bibfnamefont {J.~A.}\ \bibnamefont
  {Thornton}}, \bibinfo {author} {\bibfnamefont {K.~S.}\ \bibnamefont {Virts}},
  \bibinfo {author} {\bibfnamefont {R.~H.}\ \bibnamefont {Holzworth}}, \ and\
  \bibinfo {author} {\bibfnamefont {T.~P.}\ \bibnamefont {Mitchell}},\ }\href
  {\doibase 10.1002/2017GL074982} {\bibfield  {journal} {\bibinfo  {journal}
  {Geophysical Research Letters}\ }\textbf {\bibinfo {volume} {44}},\ \bibinfo
  {pages} {9102} (\bibinfo {year} {2017})}\BibitemShut {NoStop}%
\bibitem [{\citenamefont {Keith}(2000)}]{Keith2000}%
  \BibitemOpen
  \bibfield  {author} {\bibinfo {author} {\bibfnamefont {D.~W.}\ \bibnamefont
  {Keith}},\ }\href {\doibase 10.1146/annurev.energy.25.1.245} {\bibfield
  {journal} {\bibinfo  {journal} {Annual Review of Energy and the Environment}\
  }\textbf {\bibinfo {volume} {25}},\ \bibinfo {pages} {245} (\bibinfo {year}
  {2000})}\BibitemShut {NoStop}%
\bibitem [{\citenamefont {Keith}(2010)}]{Keith2010}%
  \BibitemOpen
  \bibfield  {author} {\bibinfo {author} {\bibfnamefont {D.~W.}\ \bibnamefont
  {Keith}},\ }\href {\doibase 10.1073/pnas.1009519107} {\bibfield  {journal}
  {\bibinfo  {journal} {Proceedings of the National Academy of Sciences}\
  }\textbf {\bibinfo {volume} {107}},\ \bibinfo {pages} {16428} (\bibinfo
  {year} {2010})}\BibitemShut {NoStop}%
\bibitem [{\citenamefont {Palmer}(1980)}]{Palmer1980}%
  \BibitemOpen
  \bibfield  {author} {\bibinfo {author} {\bibfnamefont {A.~J.}\ \bibnamefont
  {Palmer}},\ }\href {\doibase 10.1364/OL.5.000054} {\bibfield  {journal}
  {\bibinfo  {journal} {Opt. Lett.}\ }\textbf {\bibinfo {volume} {5}},\
  \bibinfo {pages} {54} (\bibinfo {year} {1980})}\BibitemShut {NoStop}%
\bibitem [{\citenamefont {Palmer}(1983)}]{Palmer1983}%
  \BibitemOpen
  \bibfield  {author} {\bibinfo {author} {\bibfnamefont {A.~J.}\ \bibnamefont
  {Palmer}},\ }\href {\doibase 10.1364/JOSA.73.001568} {\bibfield  {journal}
  {\bibinfo  {journal} {J. Opt. Soc. Am.}\ }\textbf {\bibinfo {volume} {73}},\
  \bibinfo {pages} {1568} (\bibinfo {year} {1983})}\BibitemShut {NoStop}%
\bibitem [{\citenamefont {Leung}(1985)}]{Leung1985}%
  \BibitemOpen
  \bibfield  {author} {\bibinfo {author} {\bibfnamefont {K.~M.}\ \bibnamefont
  {Leung}},\ }\href {\doibase 10.1364/OL.10.000347} {\bibfield  {journal}
  {\bibinfo  {journal} {Opt. Lett.}\ }\textbf {\bibinfo {volume} {10}},\
  \bibinfo {pages} {347} (\bibinfo {year} {1985})}\BibitemShut {NoStop}%
\bibitem [{\citenamefont {Besteiro}\ \emph {et~al.}(2017)\citenamefont
  {Besteiro}, \citenamefont {Gungor}, \citenamefont {Demir},\ and\
  \citenamefont {Govorov}}]{Besteiro2017}%
  \BibitemOpen
  \bibfield  {author} {\bibinfo {author} {\bibfnamefont {L.~V.}\ \bibnamefont
  {Besteiro}}, \bibinfo {author} {\bibfnamefont {K.}~\bibnamefont {Gungor}},
  \bibinfo {author} {\bibfnamefont {H.~V.}\ \bibnamefont {Demir}}, \ and\
  \bibinfo {author} {\bibfnamefont {A.~O.}\ \bibnamefont {Govorov}},\ }\href
  {\doibase 10.1021/acs.jpcc.6b11550} {\bibfield  {journal} {\bibinfo
  {journal} {The Journal of Physical Chemistry C}\ }\textbf {\bibinfo {volume}
  {121}},\ \bibinfo {pages} {2987} (\bibinfo {year} {2017})}\BibitemShut
  {NoStop}%
\bibitem [{\citenamefont {V.~Besteiro}\ \emph {et~al.}(2018)\citenamefont
  {V.~Besteiro}, \citenamefont {Kong}, \citenamefont {Wang}, \citenamefont
  {Rosei},\ and\ \citenamefont {Govorov}}]{Besteiro2018}%
  \BibitemOpen
  \bibfield  {author} {\bibinfo {author} {\bibfnamefont {L.}~\bibnamefont
  {V.~Besteiro}}, \bibinfo {author} {\bibfnamefont {X.-T.}\ \bibnamefont
  {Kong}}, \bibinfo {author} {\bibfnamefont {Z.}~\bibnamefont {Wang}}, \bibinfo
  {author} {\bibfnamefont {F.}~\bibnamefont {Rosei}}, \ and\ \bibinfo {author}
  {\bibfnamefont {A.~O.}\ \bibnamefont {Govorov}},\ }\href {\doibase
  10.1021/acs.nanolett.8b00764} {\bibfield  {journal} {\bibinfo  {journal}
  {Nano Letters}\ }\textbf {\bibinfo {volume} {18}},\ \bibinfo {pages} {3147}
  (\bibinfo {year} {2018})}\BibitemShut {NoStop}%
\bibitem [{\citenamefont {Bohren}\ and\ \citenamefont
  {Huffman}(1983)}]{Bohren1983}%
  \BibitemOpen
  \bibfield  {author} {\bibinfo {author} {\bibfnamefont {C.~F.}\ \bibnamefont
  {Bohren}}\ and\ \bibinfo {author} {\bibfnamefont {D.~R.}\ \bibnamefont
  {Huffman}},\ }\href@noop {} {\emph {\bibinfo {title} {Absorption and
  Scattering of Light by Small Particles}}}\ (\bibinfo  {publisher}
  {Wiley-VCH},\ \bibinfo {address} {New York, NY},\ \bibinfo {year}
  {1983})\BibitemShut {NoStop}%
\bibitem [{\citenamefont {Koman}\ \emph {et~al.}(2018)\citenamefont {Koman},
  \citenamefont {Liu}, \citenamefont {Kozawa}, \citenamefont {Liu},
  \citenamefont {Cottrill}, \citenamefont {Son}, \citenamefont {Lebron},\ and\
  \citenamefont {Strano}}]{Koman2018}%
  \BibitemOpen
  \bibfield  {author} {\bibinfo {author} {\bibfnamefont {V.~B.}\ \bibnamefont
  {Koman}}, \bibinfo {author} {\bibfnamefont {P.}~\bibnamefont {Liu}}, \bibinfo
  {author} {\bibfnamefont {D.}~\bibnamefont {Kozawa}}, \bibinfo {author}
  {\bibfnamefont {A.~T.}\ \bibnamefont {Liu}}, \bibinfo {author} {\bibfnamefont
  {A.~L.}\ \bibnamefont {Cottrill}}, \bibinfo {author} {\bibfnamefont
  {Y.}~\bibnamefont {Son}}, \bibinfo {author} {\bibfnamefont {J.~A.}\
  \bibnamefont {Lebron}}, \ and\ \bibinfo {author} {\bibfnamefont {M.~S.}\
  \bibnamefont {Strano}},\ }\href {\doibase 10.1038/s41565-018-0194-z}
  {\bibfield  {journal} {\bibinfo  {journal} {Nature Nanotechnology}\ }
  (\bibinfo {year} {2018}),\ 10.1038/s41565-018-0194-z}\BibitemShut {NoStop}%
\bibitem [{\citenamefont {Miller}\ \emph {et~al.}(2014)\citenamefont {Miller},
  \citenamefont {Hsu}, \citenamefont {Reid}, \citenamefont {Qiu}, \citenamefont
  {DeLacy}, \citenamefont {Joannopoulos}, \citenamefont {Soljacic},\ and\
  \citenamefont {Johnson}}]{Miller2014}%
  \BibitemOpen
  \bibfield  {author} {\bibinfo {author} {\bibfnamefont {O.~D.}\ \bibnamefont
  {Miller}}, \bibinfo {author} {\bibfnamefont {C.~W.}\ \bibnamefont {Hsu}},
  \bibinfo {author} {\bibfnamefont {M.~T.~H.}\ \bibnamefont {Reid}}, \bibinfo
  {author} {\bibfnamefont {W.}~\bibnamefont {Qiu}}, \bibinfo {author}
  {\bibfnamefont {B.~G.}\ \bibnamefont {DeLacy}}, \bibinfo {author}
  {\bibfnamefont {J.~D.}\ \bibnamefont {Joannopoulos}}, \bibinfo {author}
  {\bibfnamefont {M.}~\bibnamefont {Soljacic}}, \ and\ \bibinfo {author}
  {\bibfnamefont {S.~G.}\ \bibnamefont {Johnson}},\ }\href {\doibase
  10.1103/PhysRevLett.112.123903} {\bibfield  {journal} {\bibinfo  {journal}
  {Physical Review Letters}\ }\textbf {\bibinfo {volume} {112}},\ \bibinfo
  {pages} {123903} (\bibinfo {year} {2014})}\BibitemShut {NoStop}%
\bibitem [{\citenamefont {Miller}\ \emph {et~al.}(2016)\citenamefont {Miller},
  \citenamefont {Polimeridis}, \citenamefont {Homer~Reid}, \citenamefont {Hsu},
  \citenamefont {DeLacy}, \citenamefont {Joannopoulos}, \citenamefont
  {Soljacic},\ and\ \citenamefont {Johnson}}]{Miller2016}%
  \BibitemOpen
  \bibfield  {author} {\bibinfo {author} {\bibfnamefont {O.~D.}\ \bibnamefont
  {Miller}}, \bibinfo {author} {\bibfnamefont {A.~G.}\ \bibnamefont
  {Polimeridis}}, \bibinfo {author} {\bibfnamefont {M.~T.}\ \bibnamefont
  {Homer~Reid}}, \bibinfo {author} {\bibfnamefont {C.~W.}\ \bibnamefont {Hsu}},
  \bibinfo {author} {\bibfnamefont {B.~G.}\ \bibnamefont {DeLacy}}, \bibinfo
  {author} {\bibfnamefont {J.~D.}\ \bibnamefont {Joannopoulos}}, \bibinfo
  {author} {\bibfnamefont {M.}~\bibnamefont {Soljacic}}, \ and\ \bibinfo
  {author} {\bibfnamefont {S.~G.}\ \bibnamefont {Johnson}},\ }\href {\doibase
  10.1364/OE.24.003329} {\bibfield  {journal} {\bibinfo  {journal} {Optics
  Express}\ }\textbf {\bibinfo {volume} {24}},\ \bibinfo {pages} {3329}
  (\bibinfo {year} {2016})}\BibitemShut {NoStop}%
\bibitem [{\citenamefont {Kravets}\ \emph {et~al.}(2008)\citenamefont
  {Kravets}, \citenamefont {Schedin},\ and\ \citenamefont
  {Grigorenko}}]{Kravets2008}%
  \BibitemOpen
  \bibfield  {author} {\bibinfo {author} {\bibfnamefont {V.~G.}\ \bibnamefont
  {Kravets}}, \bibinfo {author} {\bibfnamefont {F.}~\bibnamefont {Schedin}}, \
  and\ \bibinfo {author} {\bibfnamefont {A.~N.}\ \bibnamefont {Grigorenko}},\
  }\href {\doibase 10.1103/PhysRevB.78.205405} {\bibfield  {journal} {\bibinfo
  {journal} {Phys. Rev. B}\ }\textbf {\bibinfo {volume} {78}},\ \bibinfo
  {pages} {205405} (\bibinfo {year} {2008})}\BibitemShut {NoStop}%
\bibitem [{\citenamefont {Park}\ \emph {et~al.}(2017)\citenamefont {Park},
  \citenamefont {Hsiao}, \citenamefont {Yi}, \citenamefont {Izor},
  \citenamefont {Koerner}, \citenamefont {Jawaid},\ and\ \citenamefont
  {Vaia}}]{Park2017}%
  \BibitemOpen
  \bibfield  {author} {\bibinfo {author} {\bibfnamefont {K.}~\bibnamefont
  {Park}}, \bibinfo {author} {\bibfnamefont {M.-s.}\ \bibnamefont {Hsiao}},
  \bibinfo {author} {\bibfnamefont {Y.-J.}\ \bibnamefont {Yi}}, \bibinfo
  {author} {\bibfnamefont {S.}~\bibnamefont {Izor}}, \bibinfo {author}
  {\bibfnamefont {H.}~\bibnamefont {Koerner}}, \bibinfo {author} {\bibfnamefont
  {A.}~\bibnamefont {Jawaid}}, \ and\ \bibinfo {author} {\bibfnamefont {R.~A.}\
  \bibnamefont {Vaia}},\ }\href {\doibase 10.1021/acsami.7b08003} {\bibfield
  {journal} {\bibinfo  {journal} {ACS Applied Materials and Interfaces}\
  }\textbf {\bibinfo {volume} {9}},\ \bibinfo {pages} {26363} (\bibinfo {year}
  {2017})}\BibitemShut {NoStop}%
\bibitem [{\citenamefont {Stoner}\ and\ \citenamefont
  {Glass}(2012)}]{Stoner2012}%
  \BibitemOpen
  \bibfield  {author} {\bibinfo {author} {\bibfnamefont {B.~R.}\ \bibnamefont
  {Stoner}}\ and\ \bibinfo {author} {\bibfnamefont {J.~T.}\ \bibnamefont
  {Glass}},\ }\href {\doibase 10.1038/nnano.2012.130} {\bibfield  {journal}
  {\bibinfo  {journal} {Nature Nanotechnology}\ }\textbf {\bibinfo {volume}
  {7}},\ \bibinfo {pages} {485} (\bibinfo {year} {2012})}\BibitemShut {NoStop}%
\bibitem [{\citenamefont {Srisonphan}\ \emph {et~al.}(2012)\citenamefont
  {Srisonphan}, \citenamefont {Jung},\ and\ \citenamefont
  {Kim}}]{Srisonphan2012}%
  \BibitemOpen
  \bibfield  {author} {\bibinfo {author} {\bibfnamefont {S.}~\bibnamefont
  {Srisonphan}}, \bibinfo {author} {\bibfnamefont {Y.~S.}\ \bibnamefont
  {Jung}}, \ and\ \bibinfo {author} {\bibfnamefont {H.~K.}\ \bibnamefont
  {Kim}},\ }\href {\doibase 10.1038/nnano.2012.107
  https://www.nature.com/articles/nnano.2012.107#supplementary-information}
  {\bibfield  {journal} {\bibinfo  {journal} {Nature Nanotechnology}\ }\textbf
  {\bibinfo {volume} {7}},\ \bibinfo {pages} {504} (\bibinfo {year}
  {2012})}\BibitemShut {NoStop}%
\bibitem [{\citenamefont {Jones}\ \emph {et~al.}(2017)\citenamefont {Jones},
  \citenamefont {Lukin},\ and\ \citenamefont {Scherer}}]{Jones2017}%
  \BibitemOpen
  \bibfield  {author} {\bibinfo {author} {\bibfnamefont {W.~M.}\ \bibnamefont
  {Jones}}, \bibinfo {author} {\bibfnamefont {D.}~\bibnamefont {Lukin}}, \ and\
  \bibinfo {author} {\bibfnamefont {A.}~\bibnamefont {Scherer}},\ }\href
  {\doibase 10.1063/1.4989677} {\bibfield  {journal} {\bibinfo  {journal}
  {Applied Physics Letters}\ }\textbf {\bibinfo {volume} {110}},\ \bibinfo
  {pages} {263101} (\bibinfo {year} {2017})}\BibitemShut {NoStop}%
\bibitem [{\citenamefont {Fontana}\ \emph {et~al.}(2016)\citenamefont
  {Fontana}, \citenamefont {da~Costa}, \citenamefont {Pereira}, \citenamefont
  {Naciri}, \citenamefont {Ratna}, \citenamefont {Palffy-Muhoray},\ and\
  \citenamefont {Carvalho}}]{Fontana2016}%
  \BibitemOpen
  \bibfield  {author} {\bibinfo {author} {\bibfnamefont {J.}~\bibnamefont
  {Fontana}}, \bibinfo {author} {\bibfnamefont {G.~K.~B.}\ \bibnamefont
  {da~Costa}}, \bibinfo {author} {\bibfnamefont {J.~M.}\ \bibnamefont
  {Pereira}}, \bibinfo {author} {\bibfnamefont {J.}~\bibnamefont {Naciri}},
  \bibinfo {author} {\bibfnamefont {B.~R.}\ \bibnamefont {Ratna}}, \bibinfo
  {author} {\bibfnamefont {P.}~\bibnamefont {Palffy-Muhoray}}, \ and\ \bibinfo
  {author} {\bibfnamefont {I.~C.~S.}\ \bibnamefont {Carvalho}},\ }\href
  {\doibase doi:http://dx.doi.org/10.1063/1.4942969} {\bibfield  {journal}
  {\bibinfo  {journal} {Applied Physics Letters}\ }\textbf {\bibinfo {volume}
  {108}},\ \bibinfo {pages} {081904} (\bibinfo {year} {2016})}\BibitemShut
  {NoStop}%
\bibitem [{\citenamefont {Etcheverry}\ \emph {et~al.}(2017)\citenamefont
  {Etcheverry}, \citenamefont {Araujo}, \citenamefont {da~Costa}, \citenamefont
  {Pereira}, \citenamefont {Camara}, \citenamefont {Naciri}, \citenamefont
  {Ratna}, \citenamefont {Hern\'{a}ndez-Romano}, \citenamefont {de~Matos},
  \citenamefont {Carvalho}, \citenamefont {Margulis},\ and\ \citenamefont
  {Fontana}}]{Etcheverry2017}%
  \BibitemOpen
  \bibfield  {author} {\bibinfo {author} {\bibfnamefont {S.}~\bibnamefont
  {Etcheverry}}, \bibinfo {author} {\bibfnamefont {L.~F.}\ \bibnamefont
  {Araujo}}, \bibinfo {author} {\bibfnamefont {G.~K.~B.}\ \bibnamefont
  {da~Costa}}, \bibinfo {author} {\bibfnamefont {J.~M.~B.}\ \bibnamefont
  {Pereira}}, \bibinfo {author} {\bibfnamefont {A.~R.}\ \bibnamefont {Camara}},
  \bibinfo {author} {\bibfnamefont {J.}~\bibnamefont {Naciri}}, \bibinfo
  {author} {\bibfnamefont {B.~R.}\ \bibnamefont {Ratna}}, \bibinfo {author}
  {\bibfnamefont {I.}~\bibnamefont {Hern\'{a}ndez-Romano}}, \bibinfo {author}
  {\bibfnamefont {C.~J.~S.}\ \bibnamefont {de~Matos}}, \bibinfo {author}
  {\bibfnamefont {I.~C.~S.}\ \bibnamefont {Carvalho}}, \bibinfo {author}
  {\bibfnamefont {W.}~\bibnamefont {Margulis}}, \ and\ \bibinfo {author}
  {\bibfnamefont {J.}~\bibnamefont {Fontana}},\ }\href {\doibase
  10.1364/OPTICA.4.000864} {\bibfield  {journal} {\bibinfo  {journal} {Optica}\
  }\textbf {\bibinfo {volume} {4}},\ \bibinfo {pages} {864} (\bibinfo {year}
  {2017})}\BibitemShut {NoStop}%
\bibitem [{\citenamefont {Gupta}\ \emph {et~al.}(2018)\citenamefont {Gupta},
  \citenamefont {Arunachalam}, \citenamefont {Cloutier},\ and\ \citenamefont
  {Izquierdo}}]{Gupta2018}%
  \BibitemOpen
  \bibfield  {author} {\bibinfo {author} {\bibfnamefont {A.~A.}\ \bibnamefont
  {Gupta}}, \bibinfo {author} {\bibfnamefont {S.}~\bibnamefont {Arunachalam}},
  \bibinfo {author} {\bibfnamefont {S.~G.}\ \bibnamefont {Cloutier}}, \ and\
  \bibinfo {author} {\bibfnamefont {R.}~\bibnamefont {Izquierdo}},\ }\href
  {\doibase 10.1021/acsphotonics.8b00829} {\bibfield  {journal} {\bibinfo
  {journal} {ACS Photonics}\ }\textbf {\bibinfo {volume} {5}},\ \bibinfo
  {pages} {3923} (\bibinfo {year} {2018})}\BibitemShut {NoStop}%
\bibitem [{\citenamefont {Phan-Quang}\ \emph {et~al.}(2018)\citenamefont
  {Phan-Quang}, \citenamefont {Lee}, \citenamefont {Teng}, \citenamefont {Koh},
  \citenamefont {Yim}, \citenamefont {Tan}, \citenamefont {Tok}, \citenamefont
  {Phang},\ and\ \citenamefont {Ling}}]{Phan2018}%
  \BibitemOpen
  \bibfield  {author} {\bibinfo {author} {\bibfnamefont {G.~C.}\ \bibnamefont
  {Phan-Quang}}, \bibinfo {author} {\bibfnamefont {H.~K.}\ \bibnamefont {Lee}},
  \bibinfo {author} {\bibfnamefont {H.~W.}\ \bibnamefont {Teng}}, \bibinfo
  {author} {\bibfnamefont {C.~S.~L.}\ \bibnamefont {Koh}}, \bibinfo {author}
  {\bibfnamefont {B.~Q.}\ \bibnamefont {Yim}}, \bibinfo {author} {\bibfnamefont
  {E.~K.~M.}\ \bibnamefont {Tan}}, \bibinfo {author} {\bibfnamefont {W.~L.}\
  \bibnamefont {Tok}}, \bibinfo {author} {\bibfnamefont {I.~Y.}\ \bibnamefont
  {Phang}}, \ and\ \bibinfo {author} {\bibfnamefont {X.~Y.}\ \bibnamefont
  {Ling}},\ }\href {\doibase 10.1002/anie.201802214} {\bibfield  {journal}
  {\bibinfo  {journal} {Angewandte Chemie International Edition}\ }\textbf
  {\bibinfo {volume} {57}},\ \bibinfo {pages} {5792} (\bibinfo {year}
  {2018})}\BibitemShut {NoStop}%
\bibitem [{\citenamefont {Raliya}\ \emph {et~al.}(2017)\citenamefont {Raliya},
  \citenamefont {Saha}, \citenamefont {Chadha}, \citenamefont {Raman},\ and\
  \citenamefont {Biswas}}]{Raliya2017}%
  \BibitemOpen
  \bibfield  {author} {\bibinfo {author} {\bibfnamefont {R.}~\bibnamefont
  {Raliya}}, \bibinfo {author} {\bibfnamefont {D.}~\bibnamefont {Saha}},
  \bibinfo {author} {\bibfnamefont {T.~S.}\ \bibnamefont {Chadha}}, \bibinfo
  {author} {\bibfnamefont {B.}~\bibnamefont {Raman}}, \ and\ \bibinfo {author}
  {\bibfnamefont {P.}~\bibnamefont {Biswas}},\ }\href {\doibase
  10.1038/srep44718
  https://www.nature.com/articles/srep44718#supplementary-information}
  {\bibfield  {journal} {\bibinfo  {journal} {Scientific Reports}\ }\textbf
  {\bibinfo {volume} {7}},\ \bibinfo {pages} {44718} (\bibinfo {year}
  {2017})}\BibitemShut {NoStop}%
\bibitem [{\citenamefont {Takenaka}\ and\ \citenamefont
  {Kitahata}(2009)}]{Takenaka2009}%
  \BibitemOpen
  \bibfield  {author} {\bibinfo {author} {\bibfnamefont {Y.}~\bibnamefont
  {Takenaka}}\ and\ \bibinfo {author} {\bibfnamefont {H.}~\bibnamefont
  {Kitahata}},\ }\href {\doibase https://doi.org/10.1016/j.cplett.2008.11.023}
  {\bibfield  {journal} {\bibinfo  {journal} {Chemical Physics Letters}\
  }\textbf {\bibinfo {volume} {467}},\ \bibinfo {pages} {327} (\bibinfo {year}
  {2009})}\BibitemShut {NoStop}%
\bibitem [{\citenamefont {Khanal}\ and\ \citenamefont
  {Zubarev}(2008)}]{Zubarev2008}%
  \BibitemOpen
  \bibfield  {author} {\bibinfo {author} {\bibfnamefont {B.~P.}\ \bibnamefont
  {Khanal}}\ and\ \bibinfo {author} {\bibfnamefont {E.~R.}\ \bibnamefont
  {Zubarev}},\ }\href {\doibase 10.1021/ja806043p} {\bibfield  {journal}
  {\bibinfo  {journal} {Journal of the American Chemical Society}\ }\textbf
  {\bibinfo {volume} {130}},\ \bibinfo {pages} {12634} (\bibinfo {year}
  {2008})}\BibitemShut {NoStop}%
\bibitem [{\citenamefont {Eustis}\ and\ \citenamefont
  {El-Sayed}(2006)}]{Eustis2006}%
  \BibitemOpen
  \bibfield  {author} {\bibinfo {author} {\bibfnamefont {S.}~\bibnamefont
  {Eustis}}\ and\ \bibinfo {author} {\bibfnamefont {M.~A.}\ \bibnamefont
  {El-Sayed}},\ }\href {\doibase 044324 10.1063/1.2244520} {\bibfield
  {journal} {\bibinfo  {journal} {Journal of Applied Physics}\ }\textbf
  {\bibinfo {volume} {100}} (\bibinfo {year} {2006}),\ 044324
  10.1063/1.2244520}\BibitemShut {NoStop}%
\bibitem [{\citenamefont {Fontana}\ \emph {et~al.}(2017)\citenamefont
  {Fontana}, \citenamefont {Nita}, \citenamefont {Charipar}, \citenamefont
  {Naciri}, \citenamefont {Park}, \citenamefont {Dunkelberger}, \citenamefont
  {Owrutsky}, \citenamefont {Pique}, \citenamefont {Vaia},\ and\ \citenamefont
  {Ratna}}]{Fontana2017}%
  \BibitemOpen
  \bibfield  {author} {\bibinfo {author} {\bibfnamefont {J.}~\bibnamefont
  {Fontana}}, \bibinfo {author} {\bibfnamefont {R.}~\bibnamefont {Nita}},
  \bibinfo {author} {\bibfnamefont {N.}~\bibnamefont {Charipar}}, \bibinfo
  {author} {\bibfnamefont {J.}~\bibnamefont {Naciri}}, \bibinfo {author}
  {\bibfnamefont {K.}~\bibnamefont {Park}}, \bibinfo {author} {\bibfnamefont
  {A.}~\bibnamefont {Dunkelberger}}, \bibinfo {author} {\bibfnamefont
  {J.}~\bibnamefont {Owrutsky}}, \bibinfo {author} {\bibfnamefont
  {A.}~\bibnamefont {Pique}}, \bibinfo {author} {\bibfnamefont
  {R.}~\bibnamefont {Vaia}}, \ and\ \bibinfo {author} {\bibfnamefont
  {B.}~\bibnamefont {Ratna}},\ }\href {\doibase 10.1002/adom.201700335}
  {\bibfield  {journal} {\bibinfo  {journal} {Advanced Optical Materials}\
  }\textbf {\bibinfo {volume} {5}},\ \bibinfo {pages} {1700335} (\bibinfo
  {year} {2017})}\BibitemShut {NoStop}%
\bibitem [{\citenamefont {Yang}\ \emph {et~al.}(2005)\citenamefont {Yang},
  \citenamefont {Wu}, \citenamefont {Wu}, \citenamefont {Wang},\ and\
  \citenamefont {Chen}}]{Yang2005}%
  \BibitemOpen
  \bibfield  {author} {\bibinfo {author} {\bibfnamefont {J.}~\bibnamefont
  {Yang}}, \bibinfo {author} {\bibfnamefont {J.~C.}\ \bibnamefont {Wu}},
  \bibinfo {author} {\bibfnamefont {Y.~C.}\ \bibnamefont {Wu}}, \bibinfo
  {author} {\bibfnamefont {J.~K.}\ \bibnamefont {Wang}}, \ and\ \bibinfo
  {author} {\bibfnamefont {C.~C.}\ \bibnamefont {Chen}},\ }\href {\doibase
  10.1016/j.cplett.2005.09.093} {\bibfield  {journal} {\bibinfo  {journal}
  {Chemical Physics Letters}\ }\textbf {\bibinfo {volume} {416}},\ \bibinfo
  {pages} {215} (\bibinfo {year} {2005})}\BibitemShut {NoStop}%
\bibitem [{\citenamefont {Neubrech}\ \emph {et~al.}(2008)\citenamefont
  {Neubrech}, \citenamefont {Pucci}, \citenamefont {Cornelius}, \citenamefont
  {Karim}, \citenamefont {Garcia-Etxarri},\ and\ \citenamefont
  {Aizpurua}}]{Neubrech2008}%
  \BibitemOpen
  \bibfield  {author} {\bibinfo {author} {\bibfnamefont {F.}~\bibnamefont
  {Neubrech}}, \bibinfo {author} {\bibfnamefont {A.}~\bibnamefont {Pucci}},
  \bibinfo {author} {\bibfnamefont {T.~W.}\ \bibnamefont {Cornelius}}, \bibinfo
  {author} {\bibfnamefont {S.}~\bibnamefont {Karim}}, \bibinfo {author}
  {\bibfnamefont {A.}~\bibnamefont {Garcia-Etxarri}}, \ and\ \bibinfo {author}
  {\bibfnamefont {J.}~\bibnamefont {Aizpurua}},\ }\href {\doibase
  10.1103/PhysRevLett.101.157403} {\bibfield  {journal} {\bibinfo  {journal}
  {Physical Review Letters}\ }\textbf {\bibinfo {volume} {101}},\ \bibinfo
  {pages} {157403} (\bibinfo {year} {2008})}\BibitemShut {NoStop}%
\bibitem [{\citenamefont {Raman}\ and\ \citenamefont {Fan}(2011)}]{Raman2011}%
  \BibitemOpen
  \bibfield  {author} {\bibinfo {author} {\bibfnamefont {A.}~\bibnamefont
  {Raman}}\ and\ \bibinfo {author} {\bibfnamefont {S.}~\bibnamefont {Fan}},\
  }\href {\doibase 10.1103/PhysRevB.83.205131} {\bibfield  {journal} {\bibinfo
  {journal} {Phys. Rev. B}\ }\textbf {\bibinfo {volume} {83}},\ \bibinfo
  {pages} {205131} (\bibinfo {year} {2011})}\BibitemShut {NoStop}%
\end{thebibliography}%


%

\end{document}